\documentclass{article}

\usepackage{spconf}
\usepackage{amsmath}
\usepackage{graphicx}
\usepackage{multicol}
\usepackage{url}
\usepackage{subcaption}

\usepackage{xcolor}
\providecommand{\edit}[1]{\textcolor{black}{{#1}}}

\title{FretNet: Continuous-Valued Pitch Contour Streaming for Polyphonic Guitar Tablature Transcription}

\name{Frank Cwitkowitz$^{1, \dagger}$\thanks{$\dagger$ Work completed as a research intern at Yousician.} \qquad Toni Hirvonen$^{2}$ \qquad Anssi Klapuri$^{2}$}
\address{$^{1}$University of Rochester, $^{2}$Yousician}

\usepackage{tikz}
\newcommand\copyrightnote{
\begin{tikzpicture}[remember picture, overlay]
\node[anchor=south, yshift=15pt] at (current page.south) {\fbox{\parbox{\dimexpr\textwidth - \fboxsep - \fboxrule\relax}
{\scriptsize Copyright 2023 IEEE. Published in ICASSP 2023 – 2023 IEEE International Conference on Acoustics, Speech and Signal Processing (ICASSP), scheduled for 4-9 June 2023 in Rhodes Island, Greece. Personal use of this material is permitted. However, permission to reprint/republish this material for advertising or promotional purposes or for creating new collective works for resale or redistribution to servers or lists, or to reuse any copyrighted component of this work in other works, must be obtained from the IEEE. Contact: Manager, Copyrights and Permissions / IEEE Service Center / 445 Hoes Lane / P.O. Box 1331 / Piscataway, NJ 08855-1331, USA. Telephone: + Intl. 908-562-3966.}}};
\end{tikzpicture}}

\begin{document}
\ninept
\maketitle
\begin{abstract}
In recent years, the task of Automatic Music Transcription (AMT), whereby various attributes of music notes are estimated from audio, has received increasing attention. At the same time, the related task of Multi-Pitch Estimation (MPE) remains a challenging but necessary component of almost all AMT approaches, even if only implicitly. In the context of AMT, pitch information is typically quantized to the nominal pitches of the Western music scale. Even in more general contexts, MPE systems typically produce pitch predictions with some degree of quantization.
In certain applications of AMT, such as Guitar Tablature Transcription (GTT), it is more meaningful to estimate continuous-valued pitch contours. Guitar tablature has the capacity to represent various playing techniques, some of which involve pitch modulation. Contemporary approaches to AMT do not adequately address pitch modulation
, and offer only less quantization at the expense of more model complexity. In this paper, we present a GTT formulation that estimates continuous-valued pitch contours, grouping them according to their string and fret of origin. We demonstrate that for this task, the proposed method significantly improves the resolution of MPE and simultaneously yields tablature estimation results competitive with baseline models.
\end{abstract}
\section{Introduction}
\label{sec:introduction}
\copyrightnote
Given a musical recording, the goal of AMT is to produce
note estimates at varying degrees of specificity. 
The task has broad applications, such as inexpensively annotating
music in the wild, providing feedback on playing in an educational setting, or searching and indexing databases based on musical content.
Typically, AMT is characterized as a combination of two sub-tasks, namely MPE and Note Tracking (NT) \cite{benetos2018automatic}. In this context, MPE is commonly formulated as the frame-level estimation of musical pitches quantized to the Western music scale 
\cite{weiss2022comparing}, and NT aims to aggregate the estimated pitch activity into predictions related to musical events, \textit{i.e.} notes \cite{kelz2019deep}.
Both MPE and NT are challenging tasks in their own right, but under this formulation the performance of both tasks can suffer. This is because the coarse resolution of MPE predictions
yields little information regarding the nuances of musical expression, and because NT is typically ill-equiped to describe \edit{pitch-varying events.}

Although the Music Information Retrieval (MIR) research community is moving toward more generalized transcription solutions \cite{gardner2021mt3, bittner2022lightweight}, these models simply cannot yet capture the expressive capacity of some instruments. In particular, the guitar is an extremely popular instrument which 
lends itself to many interesting playing techniques 
\edit{that} modulate pitch and blur the boundaries between individual notes, such as \edit{vibrato}, bends, or slides. 
Moreover, many guitarists prefer to work with tablature, a prescriptive notation capable of specifying playing techniques, when reading or annotating guitar music. In addition to MPE, the transcription of audio into tablature requires either the explicit or implicit estimation of string. Several works have addressed this problem 
\cite{wiggins2019guitar, chen2022towards, cwitkowitz2022data, kim2022note}, but these models carry the same limitations \edit{concerning discretization.}

\begin{figure}[t]
\centering
\includegraphics[width=0.815\columnwidth]{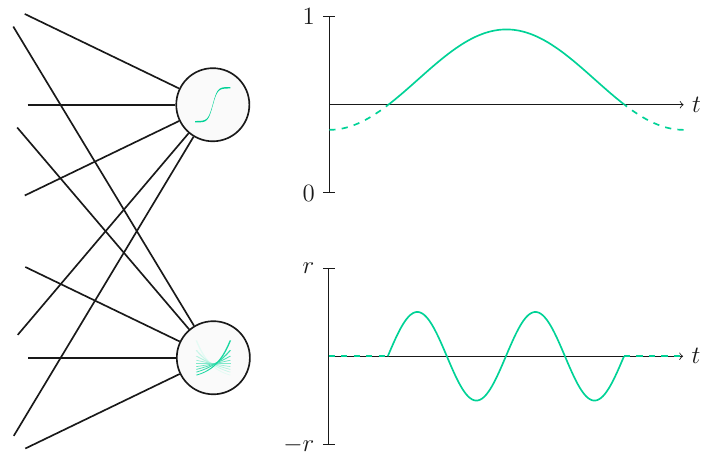}
\caption{Frame-level output targets corresponding to discrete activity and relative pitch deviation for an individual string and fret pair. (Top) sigmoid activation is applied to logits for discrete estimates. (Bottom) logits parameterize the continuous Bernoulli distribution \cite{loaiza2019continuous}, from which expected values are taken, normalized, and scaled by maximum deviation $r$ for relative pitch deviation estimates.
The two outputs combine to yield continuous-valued pitch estimates.}
\label{fig:output_structure}
\end{figure}

There is plenty of work addressing the problem of MPE under more general settings. Recent neural network based approaches \cite{kim2018crepe, singh2021deepf0} have shown much promise in this regard. However, these also produce 
pitch predictions \edit{which are quantized}, albeit to a lesser degree, and require large training datasets \edit{and increased model complexity} in order to 
\edit{estimate pitch at a higher resolution.} 
\edit{Alternatively,
some models avoid increasing complexity by estimating a pitch posteriorgram \cite{bittner2017deep, bittner2022lightweight}.}
Several \edit{pitch-tracking based} methods 
have 
been proposed to estimate playing techniques on guitar \cite{kehling2014automatic, su2019tent}, but \edit{these 
tend} to be heavily rule-based or reliant on settings with low or even no polyphony. The task of MPE is perfectly suited for the analysis of pitch modulation and various playing techniques in music, but it is still somewhat disconnected from higher-level tasks like AMT.

In this work we present \textit{FretNet}, an end-to-end GTT system capable of producing streams of continuous-valued pitch estimates, each anchored to a string and fret. \textit{FretNet} unifies the task of MPE and NT by estimating discrete activity and relative pitch deviation, achieving infinite pitch resolution in exchange for only a constant increase in model complexity.
Although we have chosen GTT as a convenient demonstration of the proposed method, our formulation is generally applicable to other tasks where the goal is to estimate continuous-valued pitch, especially within the context of music events. We conduct an ablation study to analyze the effect of the various design choices of \textit{FretNet}, and introduce metrics for evaluating note and continuous-valued pitch estimates in a string-dependent manner. Ultimately, we demonstrate that the proposed model can estimate pitch at a fraction of the resolution of contemporary models\footnote{All code is available at \url{https://github.com/cwitkowitz/guitar-transcription-continuous}.}.

\begin{table*}[t]
\begin{center}
  \begin{tabular}{|l||c|c|c||c|c|c||c|c|c||c|c|c|}
    \hline
    & \multicolumn{3}{c||}{\textbf{\edit{Tablature}}} & \multicolumn{3}{c||}{\textbf{\edit{Multi-Pitch}}} & \multicolumn{3}{c||}{\textbf{String-Dependent Note}} & \multicolumn{3}{c|}{\textbf{String-Agnostic Note}} \\
    \hline
    \multicolumn{1}{|c||}{\textbf{Experiment}} & $\mathit{P}$ & $\mathit{R}$ & $\mathit{F_1}$ & $\mathit{P}$ & $\mathit{R}$ & $\mathit{F_1}$ & $\mathit{P}$ & $\mathit{R}$ & $\mathit{F_1}$ & $\mathit{P}$ & $\mathit{R}$ & $\mathit{F_1}$ \\
    \hline
    \hline
    \textit{(1) TabCNN \cite{wiggins2019guitar}} & $0.776$ & $0.673$ & $0.717$ & $0.902$ & $\textbf{0.759}$ & $\textbf{0.820}$ & $0.398$ & $0.486$ & $0.430$ & $0.548$ & $0.656$ & $0.583$ \\
    \hline
    \textit{(2) FretNet (proposed)} & $0.801$ & $0.669$ & $0.727$ & $0.919$ & $0.742$ & $0.818$ & $0.678$ & $0.419$ & $0.506$ & $0.909$ & $0.545$ & $0.664$ \\
    \hline
    \hline
    \textit{(3) $L_{dev} \rightarrow$ MSE} & $0.803$ & $0.667$ & $0.726$ & $0.920$ & $0.738$ & $0.816$ & $\textbf{0.685}$ & $0.421$ & $0.509$ & $0.909$ & $0.542$ & $0.661$ \\
    \hline
    \textit{(4) No Deviation Head} & $0.804$ & $0.665$ & $0.726$ & $0.920$ & $0.735$ & $0.815$ & $0.682$ & $0.416$ & $0.505$ & $\textbf{0.912}$ & $0.538$ & $0.659$ \\
    \hline
    \textit{(5) No Onset Head} & $\textbf{0.805}$ & $0.665$ & $0.726$ & $\textbf{0.921}$ & $0.735$ & $0.814$ & $0.490$ & $\textbf{0.559}$ & $\textbf{0.516}$ & $0.643$ & $\textbf{0.729}$ & $\textbf{0.674}$ \\
    \hline
    \textit{(6) No Inhibition} & $0.795$ & $0.656$ & $0.717$ & $0.914$ & $0.728$ & $0.807$ & $0.674$ & $0.419$ & $0.505$ & $0.905$ & $0.545$ & $0.662$ \\
    \hline
    \textit{(7) Standard Grouping} & $0.796$ & $\textbf{0.675}$ & $\textbf{0.729}$ & $0.914$ & $0.748$ & $\textbf{0.820}$ & $0.680$ & $0.417$ & $0.505$ & $0.906$ & $0.539$ & $0.659$ \\
    \hline
    \textit{(8) CQT Features} & $0.783$ & $0.637$ & $0.700$ & $0.919$ & $0.716$ & $0.801$ & $0.647$ & $0.380$ & $0.467$ & $0.891$ & $0.508$ & $0.629$ \\
    \hline
  \end{tabular}
\end{center}
\caption{\edit{Frame-level tablature and multi-pitch results} and note-level results for string-dependent and string-agnostic criteria.
}
\label{tab:results}
\end{table*}

\section{Proposed Method}
\label{sec:proposed_method}
In this section we introduce our end-to-end GTT pipeline, which closely follows that of \cite{wiggins2019guitar} and \cite{cwitkowitz2022data}. \edit{These works introduce convolutional neural network (CNN) based models and techniques for estimating tablature at the frame-level.}
We highlight key differences as well as the novelties of our approach, and discuss our methodology for estimating continuous-valued pitch contours by string and fret.

\subsection{Feature Extraction}
\label{sec:feature_extraction}
\edit{
As input features in} \cite{wiggins2019guitar, cwitkowitz2022data}, a Constant-Q Transform (CQT) \cite{brown1991calculation} spanning 8 octaves with 2 bins per semitone and base frequency equivalent to the fundamental frequency (F0) of note \textit{C1} is computed for each piece of audio. 
While there is strong musical motivation for the CQT, the frequency support derived from this parameter setting extends beyond the F0s of a typical guitar in standard tuning. Furthermore, the resolution at higher frequencies is not sufficient to capture the energy at all relevant harmonic frequencies. Instead, we utilize the Harmonic CQT \cite{bittner2017deep}, computing multiple CQTs spanning only 4 octaves with 3 bins per semitone and base frequencies set according to the first five harmonics and the first sub-harmonic of the F0 of note \textit{E2}. 
The CQTs are stacked along a third dimension for the final feature representation. In this way, the model is encouraged to capture and exploit harmonic information organized along CQT channels during convolution \cite{bittner2017deep}. 
We resample all audio to 22050 Hz and utilize a hop size of 512 samples between frames of features.

\subsection{Model Architecture}
\label{sec:model_architecture}
The backbone 
of \textit{FretNet} is largely inspired by TabCNN \cite{wiggins2019guitar}. Perhaps the most significant changes are the deepening of the 
model and the adoption of the 
tablature prediction output layer modifications proposed in \cite{cwitkowitz2022data}. The input 
is still a 9-frame context window of features, but as a result of our modifications to feature extraction described in Sec. \ref{sec:feature_extraction}, there are \edit{six} input channels instead of \edit{one}.

\textit{FretNet} consists of three 
blocks, each comprising two 2-D convolutional layers followed by batch normalization and ReLU activation. The convolutional layers in each block utilize 3x3 kernels and contain 16, 32, and 48 filters, respectively. \edit{Temporal padding is applied in the first block.} 
After the second and third 
\edit{block}, max pooling across frequency is applied with kernel size and stride 2 and \edit{during training} dropout is applied \edit{with rates 0.5 and 0.25}, respectively. 

Embeddings are fed into three separate prediction heads. Each head consists of a fully-connected layer which reduces the dimensionality of the embeddings by half, ReLU activation, and a final fully connected layer to produce \edit{$d_{tab}$, $d_{dev}$, and $d_{ons}$} logits for 
tablature, relative pitch deviation, and onsets, respectively.
Dropout with rate 0.1 is applied before the final 
layer \edit{of} each prediction head during training.
\edit{The output sizes are $d_{tab} = 6(F + 2)$ and $d_{dev} = d_{ons} = 6(F + 1)$, where $F = 19$ is the number of frets supported.
Note that the output layer of each head 
includes neurons
for the open strings, and 
the tablature output layer includes additional neurons for the explicit modeling of string silence.}

\subsection{Estimating Continuous-Valued Pitch}
\label{sec:estimating_continuous_pitch}
The main contribution of this work is a simple and intuitive output formulation for AMT that enables a model to capture simultaneously the presence and modulation of musical events. Our design is psychoacoustically motivated in that the
human auditory system groups time-frequency activity into discrete entities (events), while being able to analyze continued and nuanced variation within those entities \cite{bregman1994auditory}. The general output structure for each string and fret pair on the guitar is illustrated in Fig. \ref{fig:output_structure}. In addition to estimating discrete activity in the same fashion as \cite{cwitkowitz2022data}, an additional output neuron produces estimates of the deviation in semitones relative to the nominal pitch associated with the respective string and fret pair.

The logits produced by the additional neurons are used to parameterize the Continuous Bernoulli distribution \cite{loaiza2019continuous}, 
which specifies the likelihood of continuous values $x \in [0, 1]$ given a single parameter. This approach was also utilized in \cite{yan2021skipping} to estimate the onsets and offsets of musical events in continuous time by predicting the relative position within frames where 
each event occurred. Although it is possible to formulate the estimation of continuous values through soft binary classification \cite{kong2021high}, this approach is incompatible with pitch deviations, which are centered around 0.5 and do not represent activations in a strict sense. 
The expected values of the parameterized distributions are computed and normalized to span $[-r, r]$, where $r$ is the 
maximum allowable pitch deviation in semitones relative to the nominal pitch of the associated string and fret pair.
Continuous-valued pitch estimates are obtained by superimposing estimated deviations onto the nominal pitch of string and fret pairs considered to be active 
using the same procedure as in \cite{cwitkowitz2022data}.

The elegance of this design is threefold. First, contemporary MPE models \cite{bittner2017deep, kim2018crepe, singh2021deepf0, bittner2022lightweight} employ an expanded output \edit{representation} to estimate discrete frequency targets with increased resolution. However, we introduce only a single neuron for each 
string and fret pair to produce continuous-valued pitch estimates. Second, by offloading the bulk of the MPE task onto the 
neurons 
\edit{tasked} with estimating pitch deviation, 
the discrete-activity neurons 
\edit{can} more appropriately
characterize musical events, \textit{i.e.} notes, 
\edit{while being less coupled to their} nominal pitches. 
The explicit pairing of neurons 
associates 
pitch estimates with the musical events 
represented by each pair. Finally, the maximum 
pitch deviation $r$ can be increased past the point where the pitch ranges of musical events 
begin to overlap. In the context of GTT, this property is useful for analyzing 
techniques such as bends, which can produce pitches several semitones higher than the nominal pitch of the corresponding string and fret.

\subsection{Event-Level Pitch Contour Streaming}
\label{sec:fret_based_contour_streaming}
In order to produce event-level guitar tablature, an onset detection head \cite{hawthorne2017onsets} is incorporated into \textit{FretNet}. The purpose of onset detection is to differentiate between sporadic and meaningful discrete activity when decoding the frame-level outputs into events. We utilize the simple decoding procedure outlined in \cite{hawthorne2017onsets} to perform this step, but do not refine any of the frame-level outputs using the final note predictions prior to evaluation.
While offsets are also generally important for AMT, 
this information is commonly left out of guitar tablature.
As such, we do not include an additional 
offset detection \edit{head} as in \cite{hawthorne2018enabling}.
Ultimately, the frame-level output of each prediction head is combined to produce note estimates with accompanying continuous-valued pitch contours. 
This information is well-suited for systems like \cite{su2019tent}, which estimate guitar playing techniques based 
\edit{on} pitch contours. Furthermore, our method is 
polyphonic, meaning it is capable of tracking multiple contours in parallel.

\subsection{Training Objectives}
\label{sec:training_objectives}
In order to balance the various objectives of all prediction heads effectively, we employ the following frame-level loss to train \textit{FretNet}:
\begin{equation}
\label{eq:objective}
L_{total} = \frac{1}{\gamma} (L_{tab} + \lambda L_{inh} + L_{ons}) + L_{dev},
\end{equation}
where $L_{tab}$, $L_{inh}$, $L_{ons}$, and $L_{dev}$ represent 
tablature, inhibition, onset, and pitch deviation loss respectively, and $\gamma$ and $\lambda$ are used as scaling parameters. 
$L_{tab}$ sums the binary cross-entropy \edit{(BCE)} loss for each string and fret pair, and $L_{inh}$ sums the product of all same-string activation pairs \cite{cwitkowitz2022data}, with $\lambda$ controlling the relative weight of 
inhibition. 
$L_{ons}$ is very similar to $L_{tab}$, but is applied to the \edit{output of the} onset detection head instead of the 
tablature 
head, with 
ground-truth activations occurring only in the first frame of each note. 
Finally, $L_{dev}$ is 
the negative log-likelihood of the ground-truth pitch deviations given the continuous Bernoulli distribution parameterized by the \edit{pitch deviation} logits. 
\edit{Note that $L_{dev}$ is not zero-bounded.}

\section{Experimental Setup}
\label{sec:experimental_setup}
In this section we detail our methodology for 
evaluating \textit{FretNet}, which follows the 
six-fold cross-validation scheme laid out in \cite{cwitkowitz2022data}.

\begin{figure}[t]
\centering
\includegraphics[width=\columnwidth]{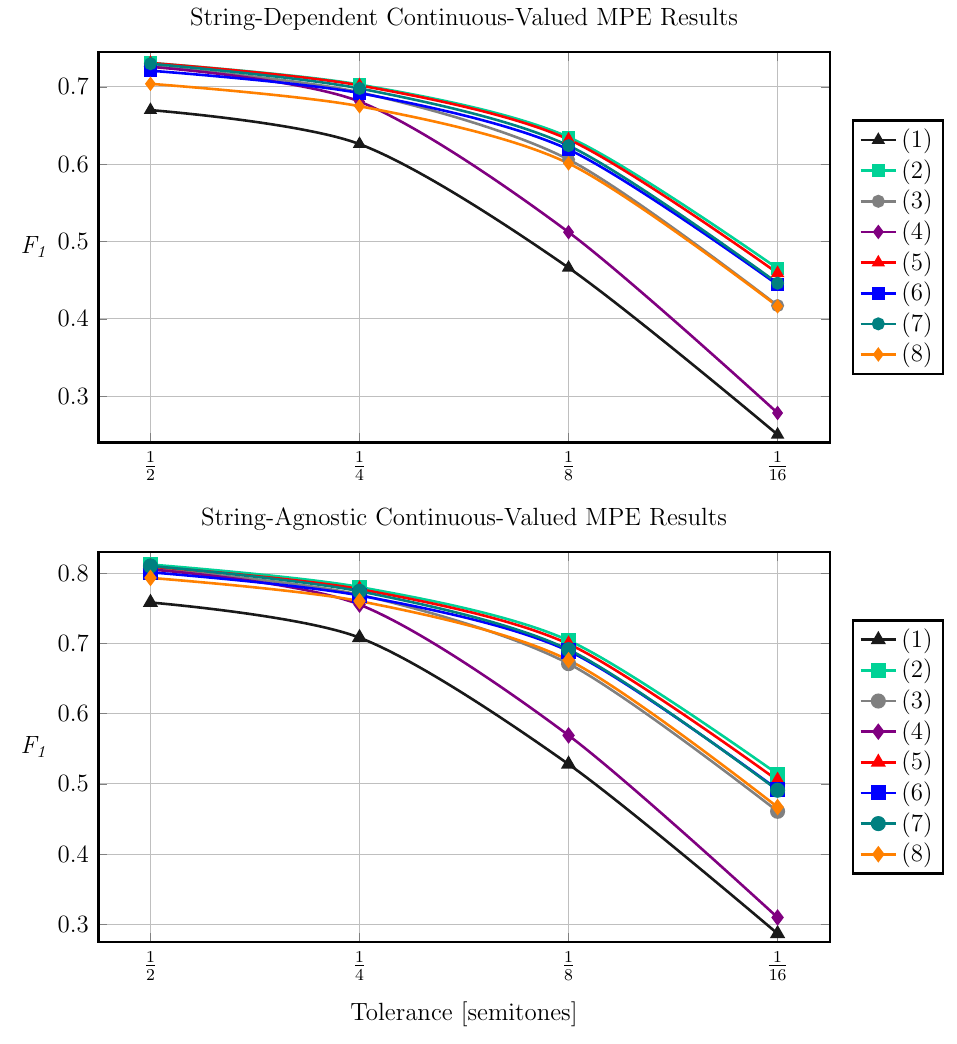}
\caption{Continuous-valued MPE results at 
\edit{various} pitch tolerances for string-dependent (top) and string-agnostic (bottom) \edit{criteria}.}
\label{fig:continuous_pitch_results}
\end{figure}

\begin{figure*}[t]
\centering
  \begin{subfigure}[t]{0.33\linewidth}
    \centering
    \includegraphics[width=1.0\linewidth]{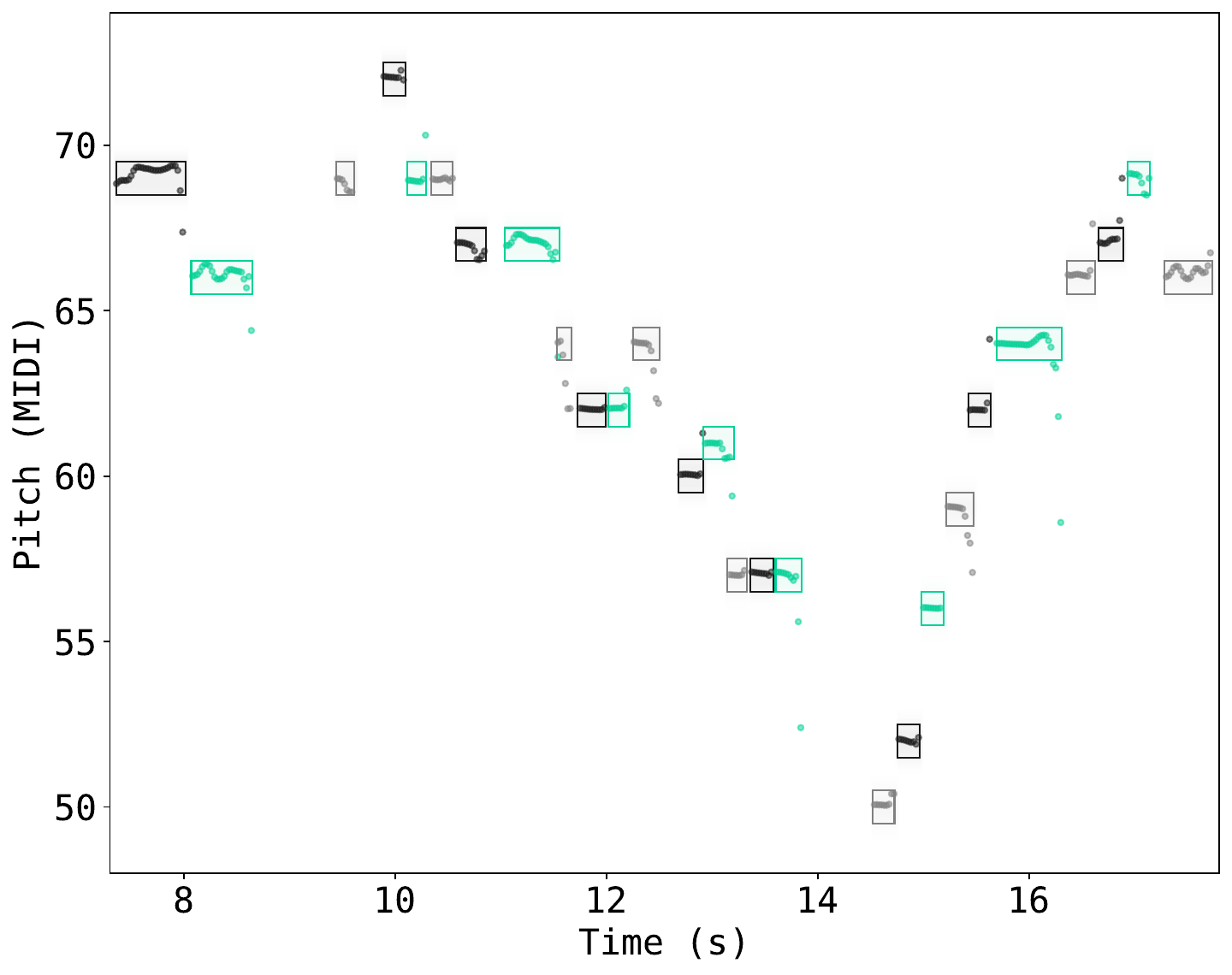}
    \caption{Ground-Truth}
  \end{subfigure}
  \hfill
  \begin{subfigure}[t]{0.33\linewidth}
    \centering
    \includegraphics[width=1.0\linewidth]{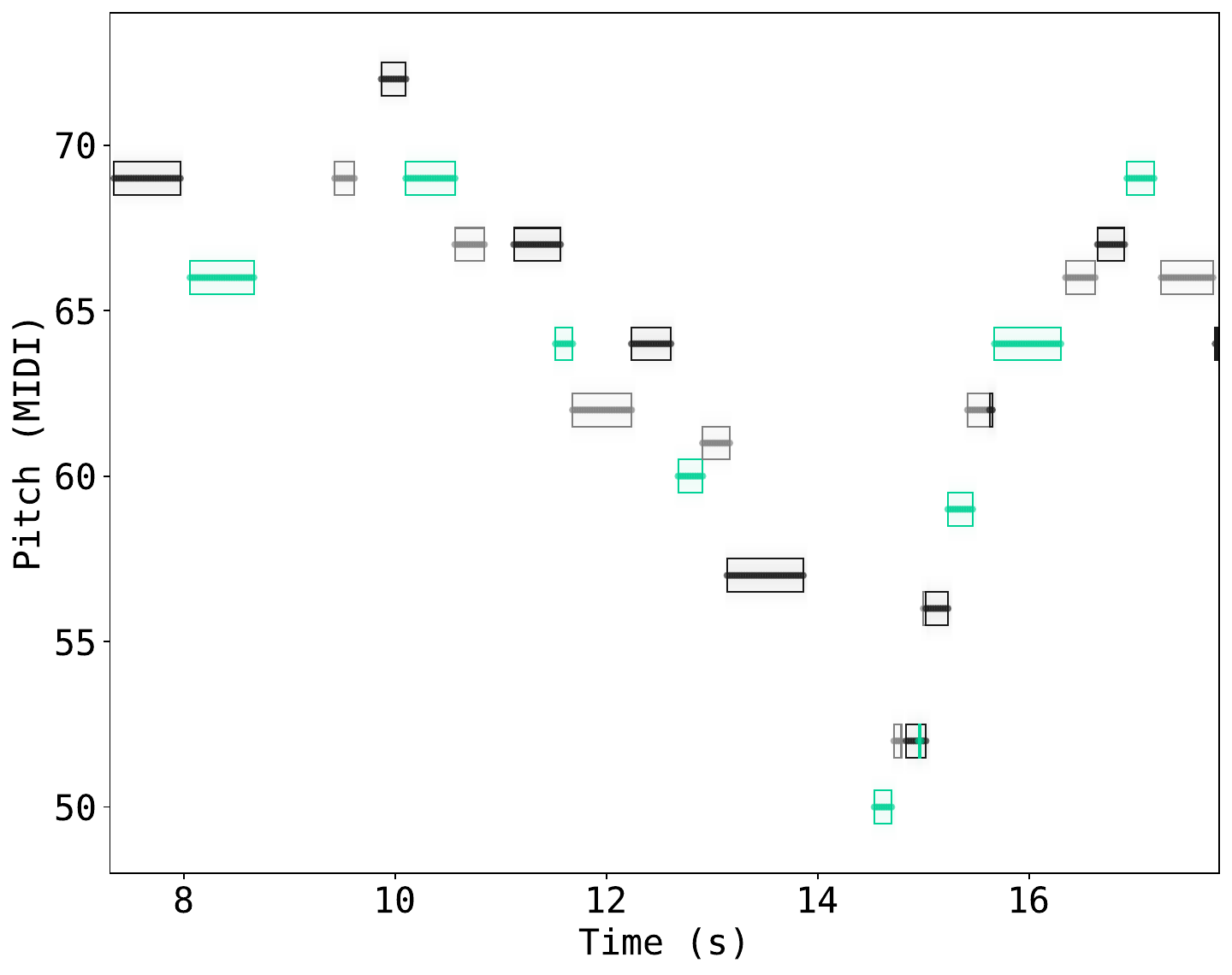}
    \caption{TabCNN Predictions}
  \end{subfigure}
  \hfill
  \begin{subfigure}[t]{0.33\linewidth}
    \centering
    \includegraphics[width=1.0\linewidth]{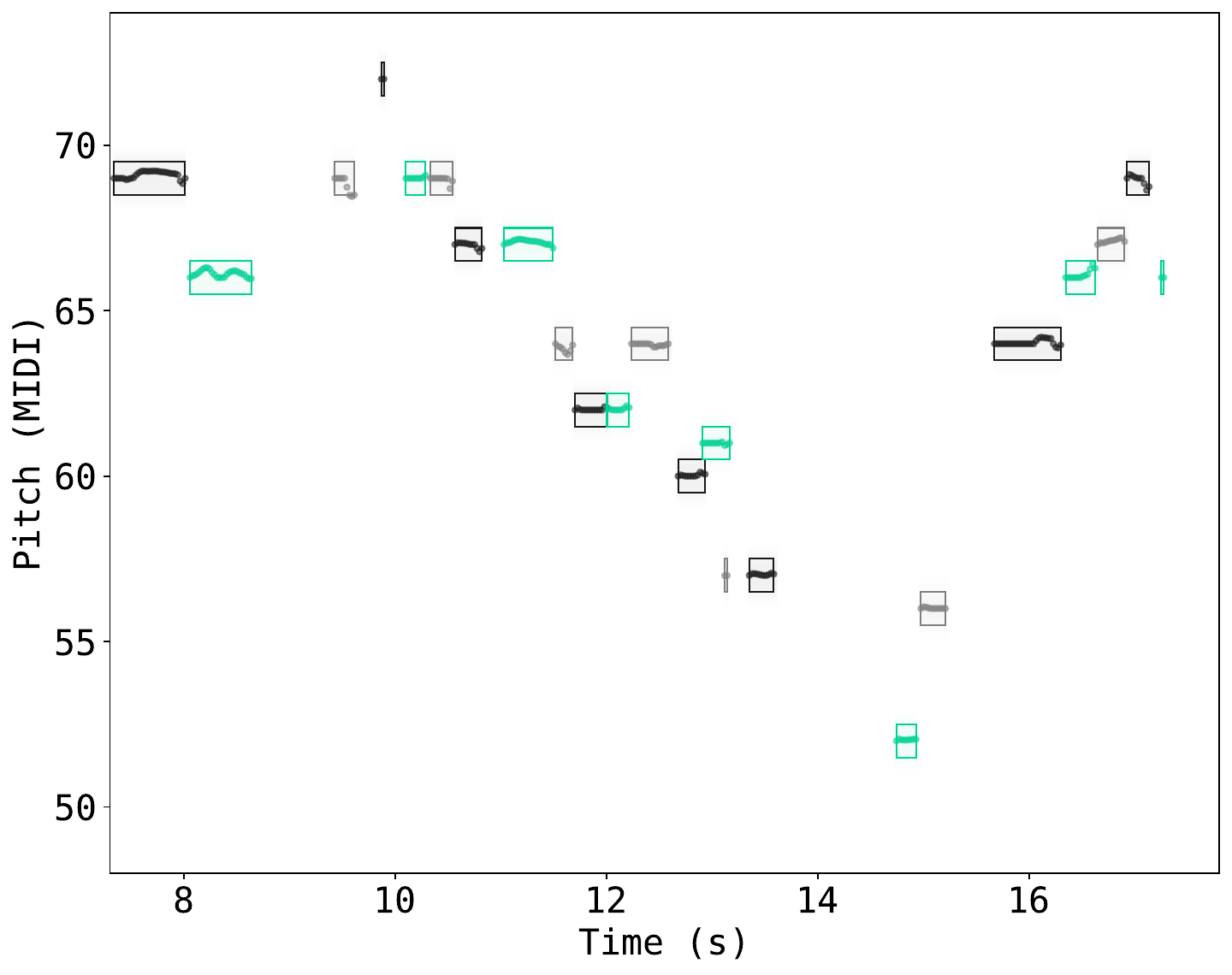}
    \caption{FretNet Predictions}
  \end{subfigure}
\caption{Comparison of ground-truth vs. predictions for track \texttt{05\_Rock1-130-A\_solo} in GuitarSet. Color indicates note-contour grouping.}
\label{fig:output_comparison}
\end{figure*}

\subsection{Dataset \& Metrics}
\label{sec:dataset}
GuitarSet \cite{quingyang2019guitarset} is a small dataset comprising audio from solo guitar playing with accompanying string-level note and pitch annotations. It consists of six players' interpretations over various chord progressions and styles, amounting in total to 360 short excerpts. We utilize GuitarSet for training, validation, and evaluation, in a six-fold cross-validation scheme split by player where 
\edit{two splits are} held out, \edit{one} for validation and evaluation, respectively. The note and pitch annotations within the dataset are grouped, meaning the string and fret origin of each pitch observation is known. This allows us to generate ground-truth targets to carry out 
continuous-valued pitch estimation 
\edit{as} described in \edit{Sec. \ref{sec:estimating_continuous_pitch}.} 
Although the note-contour grouping is provided, we implemented our own cluster-based grouping \edit{algorithm}\footnote{Please see the code for more details on this procedure.} to alleviate the effect of some noisy pitch observations (see Fig. \ref{fig:output_comparison}a). \edit{With our algorithm, pitch observations in adjacent frames with frequency difference below a certain threshold are clustered. Small clusters are discarded, and all other clusters are either assigned to the ground-truth note with the most overlap in time and nearest average frequency, 
or used to create a new note label
.}
\edit{The} 
training targets \edit{for the tablature head are} derived from \edit{discretized} note annotations \edit{and adjusted} to better align with 
\edit{the time boundaries of associated pitch contour clusters}. 
Note that 
all evaluation \edit{is performed} with respect to \edit{the} original \edit{annotations}. 
Since there is no explicit mention of playing techniques in GuitarSet, for all experiments we only utilize a maximum \edit{pitch} deviation of $r = 1.0$.

We evaluate with the original metrics proposed in \cite{wiggins2019guitar} for frame-level tablature and multi-pitch \edit{estimates. With both the inclusion (string-dependent) and exclusion (string-agnostic) of criterion for correct string,} we also compute note-level (onset only) scores and evaluate 
continuous-valued pitch predictions 
\edit{under various} pitch tolerances using \texttt{mir\_eval} \cite{raffel2014mir_eval}.
As in \cite{cwitkowitz2022data}, 
our model selection criteria for six-fold cross-validation \edit{is the frame-level tablature $F_1$-score}.

\subsection{Training Details \& Ablations}
\label{sec:training_details}
We train \textit{FretNet} with Adam optimizer for 2500 iterations using a learning rate of 0.0005, which is halved every 500 iterations. One iteration corresponds to one loop through the training partition with batch size 30, where a sequence of 200 frames, converted to context windows, is sampled from each piece. We adopt this convention 
to balance the musical statistics 
across 
pieces, irrespective of length. 
Scaling parameters $\lambda$ and $\gamma$ in Equation (\ref{eq:objective}) are both set to 10.

We establish baseline results by performing six-fold cross-validation (1) on TabCNN
\footnote{\edit{
Trained with targets derived from the original, unadjusted annotations.}}
\cite{wiggins2019guitar}, as well as (2) on \textit{FretNet} as detailed in Sec. \ref{sec:proposed_method}.
\edit{Although \textit{FretNet} adopts the output layer modifications proposed in \cite{cwitkowitz2022data}, we only employ vanilla TabCNN as a baseline, since we suspect these would not affect its note transcription or continuous-valued MPE performance.}
In order to investigate the impact of the various design choices of \textit{FretNet}, we also conduct an ablation study where we alternate one design choice at a time.

We experiment with (3) treating pitch deviation outputs as logits for sigmoid activation and training with mean squared error (MSE) instead of the continuous Bernoulli formulation.
Since MSE is zero-bounded, in this experiment we employ a modified loss function $\hat{L}_{total} = \gamma \cdot L_{total}$, such that $\gamma$ directly controls the influence of $L_{dev}$.
We experiment with (4) removing the pitch deviation 
head entirely,
in order to see if our formulation does in fact produce high-resolution pitch estimates, and also to analyze the performance floor when simply choosing the nominal pitch in each frame. We experiment with (5) removing the onset detection head entirely and directly inferring notes from clusters of 
tablature activations.
Lastly, we experiment with (6) ignoring the inhibition loss 
in Equation (\ref{eq:objective}) by setting $\lambda = 0$, (7) generating training targets from the note-contour grouping provided with GuitarSet \cite{quingyang2019guitarset}, and (8) replacing the proposed feature extraction module with the 
CQT module from 
\cite{wiggins2019guitar}.

\section{Results \& Discussion}
\label{sec:results_and_discussion}
The results of the baseline experiments and ablation study are presented in Table \ref{tab:results} and Fig. \ref{fig:continuous_pitch_results}. In terms of discrete frame-level predictions, \textit{FretNet} has comparable MPE performance to TabCNN and only slightly outperforms TabCNN in estimating 
tablature. \edit{However,} there is a large gap between the note-level performance of the models. TabCNN does not have an onset detection head, and thus note predictions can only be inferred from clusters of frame-level predictions. This actually leads to TabCNN having higher recall for note prediction, but much lower precision. Most significantly, there is an immense difference in the continuous-valued MPE performance of the two models as pitch tolerance decreases.

Most of the ablations yield comparable performance for discrete frame-level and note-level predictions, with notable exceptions being a slight decrease in overall performance without inhibition, and even further degradation when using the CQT feature extraction module. Interestingly, note prediction performance actually increases without the onset detection head. This result is surprising, but can be attributed to a sharp increase in 
note prediction recall at the expense of precision 
\edit{when an onset prediction is not required to make a note estimate}, similar to what is exhibited by TabCNN.

The best continuous-valued MPE performance is achieved by the \textit{FretNet} model with no ablations. The variations with no onset detection head, no inhibition, and the standard training targets, in that order, perform slightly worse. The variations with CQT features and the MSE formulation suffer a more significant degradation, but still clearly demonstrate an ability to perform continuous-valued pitch estimation. Unsurprisingly, the model with no \edit{pitch} deviation head performs on par with TabCNN.

We also offer a visual demonstration of predictions generated from the two baselines when presented with unseen data in Fig. \ref{fig:output_comparison}. The model checkpoints 
were chosen from the fifth fold of the respective experiments using the selection criteria defined in Sec. \ref{sec:dataset}. \textit{FretNet} is able to generate continuous-valued pitch contours grouped by note, 
\edit{and we} observe that \edit{noisy pitches are not carried over} 
from the ground-truth, likely due to our cluster-based note-contour grouping 
\edit{which} discards sporadic pitches. It is also evident that TabCNN repeats several note predictions on multiple strings, further explaining 
\edit{its} increased recall and lower precision for note-level \edit{estimates.}

\section{Conclusion}
\label{sec:conclusion}
In this work, we have presented a unified model and methodology for estimating continuous-valued pitch contours within the context of guitar tablature transcription. 
Our experiments indicate that the proposed model is able to produce pitch estimates at a much higher resolution than contemporary models,
without incurring any degradation with respect to 
other integral tasks. We believe our work \edit{sheds light on} 
a promising direction for the holistic analysis of musical performances and 
playing techniques, 
and that future work should further investigate the task of continuous-valued pitch estimation and apply it more broadly to other instruments and applications.

\pagebreak
\onecolumn
\begin{multicols}{2}
\bibliographystyle{IEEEbib}
\bibliography{refs}

\begin{thebibliography}{10}

\bibitem{benetos2018automatic}
Emmanouil Benetos, Simon Dixon, Zhiyao Duan, and Sebastian Ewert,
\newblock ``Automatic music transcription: An overview,''
\newblock {\em IEEE Signal Processing Magazine}, vol. 36, no. 1, pp. 20--30,
  2019.

\bibitem{weiss2022comparing}
Christof Wei{\ss} and Geoffroy Peeters,
\newblock ``Comparing deep models and evaluation strategies for multi-pitch
  estimation in music recordings,''
\newblock {\em IEEE/ACM Transactions on Audio, Speech, and Language Processing
  (TASLP)}, vol. 30, pp. 2814--2827, 2022.

\bibitem{kelz2019deep}
Rainer Kelz, Sebastian Böck, and Gerhard Widmer,
\newblock ``Deep polyphonic {ADSR} piano note transcription,''
\newblock in {\em Proceedings of ICASSP}, 2019.

\bibitem{gardner2021mt3}
Josh Gardner, Ian Simon, Ethan Manilow, Curtis Hawthorne, and Jesse Engel,
\newblock ``{MT3}: Multi-task multitrack music transcription,''
\newblock in {\em Proceedings of ICLR}, 2021.

\bibitem{bittner2022lightweight}
Rachel~M. Bittner, Juan~Jos{\'e} Bosch, David Rubinstein, Gabriel
  Meseguer-Brocal, and Sebastian Ewert,
\newblock ``A lightweight instrument-agnostic model for polyphonic note
  transcription and multipitch estimation,''
\newblock in {\em Proceedings of ICASSP}, 2022.

\bibitem{wiggins2019guitar}
Andrew Wiggins and Youngmoo Kim,
\newblock ``Guitar tablature estimation with a convolutional neural network,''
\newblock in {\em Proceedings of ISMIR}, 2019.

\bibitem{chen2022towards}
Yu-Hua Chen, Wen-Yi Hsiao, Tsu-Kuang Hsieh, Jyh-Shing~Roger Jang, and Yi-Hsuan
  Yang,
\newblock ``Towards automatic transcription of polyphonic electric guitar
  music: A new dataset and a multi-loss transformer model,''
\newblock in {\em Proceedings of ICASSP}, 2022.

\bibitem{cwitkowitz2022data}
Frank Cwitkowitz, Jonathan Driedger, and Zhiyao Duan,
\newblock ``A data-driven methodology for considering feasibility and pairwise
  likelihood in deep learning based guitar tablature transcription systems,''
\newblock in {\em Proceedings of SMC}, 2022.

\bibitem{kim2022note}
Sehun Kim, Tomoki Hayashi, and Tomoki Toda,
\newblock ``Note-level automatic guitar transcription using attention
  mechanism,''
\newblock in {\em Proceedings of EUSIPCO}, 2022.

\bibitem{loaiza2019continuous}
Gabriel Loaiza-Ganem and John~P. Cunningham,
\newblock ``The continuous bernoulli: Fixing a pervasive error in variational
  autoencoders,''
\newblock in {\em Proceedings of NeurIPS}, 2019.

\bibitem{kim2018crepe}
Jong~Wook Kim, Justin Salamon, Peter Li, and Juan~P. Bello,
\newblock ``Crepe: A convolutional representation for pitch estimation,''
\newblock in {\em Proceedings of ICASSP}, 2018.

\bibitem{singh2021deepf0}
Satwinder Singh, Ruili Wang, and Yuanhang Qiu,
\newblock ``{DeepF0}: End-to-end fundamental frequency estimation for music and
  speech signals,''
\newblock in {\em Proceedings of ICASSP}, 2021.

\bibitem{bittner2017deep}
Rachel~M. Bittner, Brian McFee, Justin Salamon, Peter Li, and Juan~P. Bello,
\newblock ``Deep salience representations for f0 estimation in polyphonic
  music,''
\newblock in {\em Proceedings of ISMIR}, 2017.

\bibitem{kehling2014automatic}
Christian Kehling, Jakob Abe{\ss}er, Christian Dittmar, and Gerald Schuller,
\newblock ``Automatic tablature transcription of electric guitar recordings by
  estimation of score and instrument-related parameters,''
\newblock in {\em Proceedings of DAFx}, 2014.

\bibitem{su2019tent}
Ting-Wei Su, Yuan-Ping Chen, Li~Su, and Yi-Hsuan Yang,
\newblock ``Tent: Technique-embedded note tracking for real-world guitar solo
  recordings,''
\newblock {\em Transactions of the International Society for Music Information
  Retrieval (TISMIR)}, vol. 2, no. 1, pp. 15--28, 2019.

\bibitem{brown1991calculation}
Judith~C. Brown,
\newblock ``Calculation of a constant q spectral transform,''
\newblock {\em Journal of the Acoustical Society of America (JASA)}, vol. 89,
  no. 1, pp. 425--434, 1991.

\bibitem{bregman1994auditory}
Albert~S. Bregman,
\newblock {\em Auditory Scene Analysis: The Perceptual Organization of Sound},
\newblock MIT Press, 1994.

\bibitem{yan2021skipping}
Yujia Yan, Frank Cwitkowitz, and Zhiyao Duan,
\newblock ``Skipping the frame-level: Event-based piano transcription with
  neural semi-{CRF}s,''
\newblock in {\em Proceedings of NeurIPS}, 2021.

\bibitem{kong2021high}
Qiuqiang Kong, Bochen Li, Xuchen Song, Yuan Wan, and Yuxuan Wang,
\newblock ``High-resolution piano transcription with pedals by regressing onset
  and offset times,''
\newblock {\em IEEE/ACM Transactions on Audio, Speech, and Language Processing
  (TASLP)}, vol. 29, pp. 3707--3717, 2021.

\bibitem{hawthorne2017onsets}
Curtis Hawthorne, Erich Elsen, Jialin Song, Adam Roberts, Ian Simon, Colin
  Raffel, Jesse Engel, Sageev Oore, and Douglas Eck,
\newblock ``Onsets and frames: Dual-objective piano transcription,''
\newblock in {\em Proceedings of ISMIR}, 2018.

\bibitem{hawthorne2018enabling}
Curtis Hawthorne, Andriy Stasyuk, Adam Roberts, Ian Simon, Cheng-Zhi~Anna
  Huang, Sander Dieleman, Erich Elsen, Jesse Engel, and Douglas Eck,
\newblock ``Enabling factorized piano music modeling and generation with the
  {MAESTRO} dataset,''
\newblock in {\em Proceedings of ICLR}, 2019.

\bibitem{quingyang2019guitarset}
Qingyang Xi, Rachel~M. Bittner, Johan Pauwels, Xuzhou Ye, and Juan~P. Bello,
\newblock ``{GuitarSet}: A dataset for guitar transcription,''
\newblock in {\em Proceedings of ISMIR}, 2018.

\bibitem{raffel2014mir_eval}
Colin Raffel, Brian McFee, Eric~J. Humphrey, Justin Salamon, Oriol Nieto, Dawen
  Liang, and Daniel~P.W. Ellis,
\newblock ``mir\_eval: A transparent implementation of common {MIR} metrics,''
\newblock in {\em Proceedings of ISMIR}, 2014.

\end{thebibliography}
\end{multicols}

\end{document}